\documentclass[aps,prb,groupedaddress,twocolumn,10pt]{revtex4-1}

\begin{document}

\newcommand{\be}{\begin{equation}}
\newcommand{\ee}{\end{equation}}
\newcommand{\bea}{\begin{eqnarray}}
\newcommand{\eea}{\end{eqnarray}}

\title{Die Hard Holographic Phenomenology of Cuprates}

\author{D. V. Khveshchenko}
\affiliation{Department of Physics and Astronomy, 
University of North Carolina, Chapel Hill, NC 27599}

\begin{abstract}
\noindent
This note discusses the attempts of fitting a number of the approximate
power-law dependencies observed in the cuprates  
into one consistent holographic or holographically inspired hydrodynamic framework. Contrary to the expectations, the goal of reproducing as many as possible of the established behaviors of the
thermodynamic and transport coefficients appears to be achievable within the simplistic picture of a non-degenerate fermion fluid with 
quadratic dispersion. While not immediately elucidating the essential physics 
of the cuprates, this observation suggests a possible reason for
which the previous attempts towards that goal have remained inconclusive.  
  
\noindent
\end{abstract}
\maketitle

\noindent
{\it Transport in cuprates}\\

The normal state of the cuprate superconductors has long remained a challenge 
defying many attempts of its theoretical understanding. 
In the continuing absence of a fully satisfactory microscopic description, a modest goal 
has been that of constructing a more or less successful phenomenological description capable of accounting 
for most of the observed transport properties. 

Initially, the phenomenologies of the cuprates focused on the much publicized 
dichotomy between the robust power-law behaviors of the longitudinal conductivity
observed in the optimally doped $YBCO$ (and, to a lesser extent, $LSCO$) compounds
\be
\sigma\sim T^{\alpha}
\ee
with $\alpha_{exp}=-1$ and the Hall angle
\be
\tan\theta_H\sim T^{\beta}
\ee
manifesting the exponent $\beta_{exp}=-2$.

In the early theoretical proposals, Eqs.(1,2) were argued to imply the existence of two distinct 
scattering times: $\tau\sim T^{-1}$ and  $\tau_{H}\sim T^{-2}$ which were supposed to 
characterize the relaxation of either longitudinal vs transverse \cite{pwa1}, 
charge-symmetric vs anti-symmetric \cite{coleman} currents, or a 
two-fluid nature of charge and heat transport \cite{palee}.
Yet another insightful proposal of the 'marginal Fermi liquid' 
phenomenology was put forward early on \cite{varma}. 

Additional evidence of anomalous transport in the cuprates was provided by the magnetoresistivity 
\be
{\Delta\rho\over \rho}\sim T^{\gamma}
\ee
that violates the conventional Kohler's law $\Delta\rho/\rho\sim B^2/\rho^2$, instead featuring the 
exponent $\gamma_{exp}=-4$ \cite{harris}
(in strong fields $B\gg T$ the quadratic field dependence changes to a linear one).

The anomalous transport properties () appear to co-exist with 
the fairly conventional thermodynamic ones, including the Fermi-liquid-like specific heat and entropy   
\be
c\sim s\sim T^{\nu}
\ee
with $\nu_{exp}=1$ \cite{loram}. In some compounds, upon approaching 
the pseudogap phase $c(T)$ can also show a logarithmic enhancement, possibly 
signifying a quantum phase transition \cite{michon1}.

In the presence of thermal gradients, the combined thermo-electric response is
described by the coefficients relating the charge $\bf J$  
and heat $\bf Q$ currents to the gradients of electric potential and temperature
\bea
{\bf J}={\hat \sigma} {\bf E}-{\hat \alpha} {\bf \nabla}T\nonumber\\
{\bf Q}=T{\hat \alpha} {\bf E}-{\hat {\bar \kappa}} {\bf \nabla}T
\eea
where the $2\times 2$ matrices such as, e.g.,   
${\hat \sigma}_{ij}=\sigma {\delta}_{ij}+\sigma_H {\epsilon}_{ij}$,
are composed of the longitudinal and transverse (Hall) components.   

Early on the studies of heat transport focused on the Hall component as 
its longitudinal counterpart is believed to be dominated by the phonon contribution
in most of the phase diagram \cite{ong}. 
However, in the presence of chiral spin structures a sizable $\kappa_H$ signal might also stem from phonons or magnons, or even both \cite{griss1}.

The list of the actually measured observables includes  
thermopower (Seebeck) coefficient, thermal conductivities at zero current,
the Hall Lorenz number, and Nernst coefficient
\bea
S=\alpha/\sigma,~~~~~ 
{\kappa}={\bar \kappa}-\alpha^2/\sigma \nonumber\\ 
L_H=\kappa_H/T\sigma_H,~~~~~
e_N={\alpha_H\sigma-\alpha\sigma_H\over \sigma^2+\sigma^2_H}
\eea
For some of these quantities the available data still remain scarce and their independent verification is badly needed.  
Nonetheless, the above coefficients (in the case of thermopower, its deviation from 
a possible constant term) might also exhibit the power-law dependencies
\bea
\kappa_H\sim T^{\delta},~~~~~ 
L_H\sim T^{\lambda},\nonumber\\
e_N\sim T^{\mu},~~~~~
S\sim T^{\rho}~~~~~~
\eea
where $\rho_{exp}\approx 1$, $\delta_{exp},\mu_{exp}<0$, and $\lambda_{exp}\geq 0$.
   
More specifically, in Ref.\cite{ong} the data on $L_H$ in the untwinned samples of
optimally doped $YBaCuO$ were fitted into a linear dependence ($\lambda=1$) 
while the Nernst signal $e_N$ was found to increase dramatically with decreasing temperature. This effect 
was attributed to the superconducting fluctuations and/or disordered vortex pairs whose 
(positive) contribution dominates over that of the quasiparticles 
(whose sign, in turn, depends on the dominant type of carriers) upon approaching $T_c$.
Besides, $e_N$ turned out to be strongly affected by a proximity to the 
pseudogap regime and can even become anisotropic \cite{chang}.

However, the subsequent Ref.\cite{matusiak} reported somewhat different results for $\sigma_H$ and $\kappa_H$.  
and the concomitant slower temperature dependence of $L_H$ in the $LaSrCuO$, $EuBaCuO$, and  
$YBaCuO$ compounds. Specifically, in the twinned $YBaCuO$ samples the measured exponents were 
\be 
\beta^{\cite{matusiak}}_{exp}=-1.7,~~~~~\delta^{\cite{matusiak}}_{exp}=-1.2,
~~~~~\lambda^{\cite{matusiak}}_{exp}=0.5
\ee
Unlike Ref.\cite{ong}, the measurements in Ref.\cite{matusiak} of, both, $\sigma_H$ and $\kappa_H$ were 
carried out on the same, rather than different, samples. 

Adding to the puzzle of the cuprates' transport properties, there have been persistent  
reports of the Fermi liquid-like rate of inelastic quasiparticle scattering \cite{barisic}
\be
\Gamma_{qp}\sim T^{2}
\ee
in contrast to the almost uniformly accepted \cite{zaanen1} and seemingly ubiquitous 
(see, however, Ref.\cite{sad}) 'Planckian' dissipation rate that is believed to control 
local equilibration/thermalization
\be
\Gamma_{eq}\sim T
\ee
Generally, the latter would be expected 
in a quantum-critical phase associated with a quantum phase transition  
and in the absence of an intrinsic energy scale, other than temperature.

In the context of the cuprates, a number of the potentially
viable quantum critical transitions have been discussed, their list including superconducting, 
spin, charge, nematic, as well as other, even more exotic, instabilities. However, some 
data \cite{pelc} might indicate that the quantum critical scenario may not necessarily be at work.  

However, the staunch belief in the universality of Eq.(10) and its interpretation as a key 
evidence in support of the strong (as opposed to just moderate) correlations in the cuprates has brought 
to life a number of proposals based on the various 'ad hoc' generalizations of the original 
ground-breaking conjecture of holographic correspondence. \\

\noindent
{\it Applied holography}\\

In its own words, the 'bottom-up' applied holography (a.k.a. $AdS/CMT$ or Anti-de-Sitter/Condensed matter theory
correspondence) purports to offer a unique, intrinsically strong-coupling, 
approach to a variety of the traditionally 
hard condensed matter problems \cite{ads}. On the technical side, 
this intriguing (albeit still lacking a solid proof) scheme borrows 
 its computational apparatus (in essence, 'ad verbatim') from the original machinery of the conjectured 
holographic $AdS/CFT$ (Anti-de-Sitter/Conformal Field Theory) correspondence which was 
developed and professed in the 'bona fide' string/field theory. 

From the conceptual standpoint, searching for a common cause of the observed properties 
would indeed make perfect sense if the sought-after universality were indeed present. 
However, under a closer inspection  
even some close members of the same family of materials often demonstrate different behaviors and 
exhibit different power-laws. Obviously, any significant diversity between  
the related compounds would be rather difficult to accommodate under the holographic paradigm, since 
virtually every compound would then require individual treatment and a material-specific dual bulk geometry. 

Such potential difficulties notwithstanding, 
the decade-long vigorous work on the $AdS/CMT$ opportunistically 
explored a variety of the popular geometries
(Reissner-Nordstrom, Lifshitz, hyperscaling-violating, Bianchi, Q-lattices, etc.) \cite{ads}. 
Such exploratory studies resulted in a number of rather exotic proposals for obtaining 
some of the exponents (1-4,7), although in order to reproduce even 
the basic Eqs.(1,2) such analyses would often go to quite a length.  

For example, one of the popular schemes \cite{zaanen2} 
invokes the extreme 'ultra-local' $AdS_2$ limit where, both, the dynamical 
critical index $z$ and the 'hyperscaling-violation' exponent $\theta$ take 
infinite values, thereby conspiring to make 
the conductivity inversely proportional to the entropy density  
\be
\sigma\sim T^{(\theta-2-z)/z}\sim s^{-1}T^{-2/z}
\ee  
in order to conform to Eq.(1) for $z\to\infty$.

Another, more comprehensive, attempt was made in Ref.\cite{hk} where
the values $z=4/3$ and $\theta=0$ were argued to reproduce the observed behavior 
of the transport coefficients (1-3) while $\lambda^{\cite{hk}}_{holo}=1$ was chosen as one of the 
constitutive relations (despite  the fact that, unlike Ref.\cite{ong},
the works of Refs.\cite{matusiak,chang} reported
a slower-than-linear in - or even decreasing with - $T$ electronic Lorenz ratio), yet still other exponents 
\be 
\nu^{\cite{hk}}_{holo}=1.5, ~~~~~\rho^{\cite{hk}}_{holo}=0.5, ~~~~~\mu^{\cite{hk}}_{holo}=-1.5
\ee 
were markedly off their targeted values (4) and
\be 
\rho_{exp}=0~~~ or~~~ 1 
\ee
depending on whether the goal was to fit the constant (which may or may not have been of electronic origin)
or the linear term in the experimental plot for $LaSrCuO$ \cite{kim}
\be 
S=a-bT
\ee
In that regard, the analysis of Ref.\cite{hk} ignored the constant and went straight for the $T$-dependent term.  
 
For the sake of completeness and as one example of an alternate scaling scheme characterized by the least exotic 
values $z=\theta=1$ it might be worth mentioning the little known (let alone, cited) Ref.\cite{dvk1} where, 
the exponents Eqs.(1-4), alongside 
\be 
\nu^{\cite{dvk1}}_{holo}=1,~~~~~\rho^{\cite{dvk1}}_{holo}=1
\ee
and, to a lesser extent, $\lambda^{\cite{dvk1}}_{holo}=0$ 
were found to generally agree with experiment (there was not enough data available for ascertaining 
the predicted values of $\mu^{\cite{dvk1}}_{holo}=-1$ and $\delta^{\cite{dvk1}}_{holo}=-2$, though). 

Furthermore, the results of Ref.\cite{hk} were argued to compare 
favorably with certain characteristics of the energy- and momentum-dependent 
magnetic susceptibility, as probed by inelastic neutron scattering in $LaSrCuO$.
Specifically, the predicted behavior of the ratio (here $\bf Q$ is the antiferromagnetic
vector)      
\be
{\chi_s(\omega,{\bf q})\over \omega}|_{\omega\to 0}\sim |{\bf q}-{\bf Q}|^{\eta}
\ee
was found to be governed by the exponent $\eta_{holo}=-10/3$ which was indeed close to the measured $\eta_{exp}=-3$ \cite{mook}. Curiously, though, the alternate scheme
of Ref.\cite{dvk1} yielded the exponent $\eta=-3$ which was right on the data.

Also, the momentum integral 
$
T\int d{\bf q}{\chi_s(\omega,{\bf q})\over \omega}|_{\omega\to 0}
$
turned out to be constant in both schemes of Refs.\cite{hk} and \cite{dvk1}, 
again in agreement with the data of Ref.\cite{mook}.

In turn, the uniform magnetic  and charge susceptibilities 
\be   
\chi_s={d^2f\over dB^2}\sim T^{\xi},~~~~~
\chi_c={d^2f\over d\mu^2}\sim T^{\zeta}
\ee
were found to be governed by the exponents
\be 
\xi^{\cite{hk}}_{holo}=-1.5,~~~~~ \zeta^{\cite{hk}}_{holo}=0.5
\ee
 and 
\be \xi^{\cite{dvk1}}_{holo}=-2, ~~~~~\zeta^{\cite{dvk1}}_{holo}=0
\ee 
in Refs.\cite{hk} and \cite{dvk1}, respectively, 
thus providing additional means of discriminating between the two scenaria.
Notably, in either scheme the Wilson ratio $c/\chi_cT$ conformed to a constant, again 
in accord with experiment. 

Although a couple of subsequent publications \cite{ge} 
acknowledged (somewhat reluctantly) the somewhat better agreement 
with the mundane predictions of Ref.\cite{dvk1}, 
great many other (remarkably look-alike and customarily verbose) 
holographic papers pursued these and related topics relentlessly, as if
their sheer number and volume \cite{search} were to live up to  
the celebrated 'more is different' \cite{pwa2} principle.

However, being often published in such (rather uncommon for condensed matter) venues 
as $JHEP$ or $Phys.Rev.D$ quite a few of those works seem to have escaped the attention of 
(and, incidentally, avoided a closer scrutiny by) the traditional condensed matter community. 

Recently, though, the number of papers on the topic of $AdS/CMT$ has markedly 
decreased from what once seemed like an endless flurry down to a mere trickle.
Conceivably, this was a reflection of the fact that, despite much of the initial enthusiasm and effort, 
all the previous attempts of putting the holographic phenomenology on a firm 
foundation (either along the lines of the geometrized RG flow or entanglement 
dynamics in tensor networks, or by using artificial thermodynamic/information (Fisher-Ruppeiner, 
Fubini-Study, etc.) metrics, or else) have so far remained consistently inconclusive. 

In that regard, the currently popular solvable low-dimensional examples of holographic correspondence involving   
the $0+1\to 1+1$-dimensional ones (e.g., $SYK/JT$) and their $1+1\to 2+1$ counterparts (e.g., $KdV/BTZ$) can not be 
viewed as genuinely holographic. Indeed, the gravitational sectors of their bulk duals 
($1+1$- and $2+1$- dimensional, respectively) 
are non-dynamical and fully determined by their boundary (that is, $0+1$ and $1+1$-dimensional) 
degrees of freedom, thus merely revealing some intricate connections 
between the different realizations of the conformal (Virasoro) group 
(chiral in $0+1$ and doubled non-chiral in $1+1$ dimensions, respectively) and 
the different variants of its co-adjoint orbit quantization. 

That said, it might be, of course, still possible for some form of generalized holography to be derivable from a certain
fundamental principle, thereby making its original string-theoretical connection largely historic and unnecessary \cite{dvk2}. 

To that end, in the meantime the practical $AdS/CMT$ has been reinventing  
itself as advanced hydrodynamics of strongly coupled quantum matter.
Correspondingly, instead of the once ubiquitous futuristic pictures of 
esoteric black holes, nowadays a typical presentation on the topic of $AdS/CMT$ is more likely to 
feature water flows, whirlpools, and other Earthly hydrodynamic patterns \cite{holotube}.  

The renewed appreciation for and novel applications of such a well established 
field as hydrodynamics  (which, while suggesting some formal holographic connections, had long been 
discussed before the rise of holography) emerged out of the recent experimental discoveries of 
the electron hydrodynamic regime in mono- and bi-layer graphene, $(Al,Ga)As$ heterostructures, $PdCoO_2$, Herbertsmithite, etc. 
Among other things, it also resulted in a renewed interest in the anomalous transport in the cuprates.\\

\noindent
{\it (Non)holographic Hydrodynamics} \\

The intimate relation between classical gravity and hydrodynamics has 
long been known as a particular take on the $AdS/CFT$ referred to as 'fluid-gravity' correspondence \cite{ads}.  The crux of the matter lies in the deep similarity between the asymptotic near-boundary behavior of the Einstein equations for the bulk metric and the Navier-Stokes ones describing a dual boundary 
fluid in one lesser dimension. Albeit usually truncated and, therefore, approximate 
such relations can be systematically improved, thus enabling certain computational simplifications.   

The magneto-hydrodynamic transport coefficients were first derived in the early work of Ref.\cite{hydro} under the assumption of a (pseudo)relativistic 
kinematics of the charge and heat carriers. While the underlying hydrodynamic 
equations were mimicked after those of a quark-gluon plasma, they would also be considered applicable to the electron transport in graphene (the actual hydrodynamic equations describing mono-layer graphene appear to be somewhat different due to the presence of an extra hydrodynamic 'imbalance' mode 
as well as the expressly non-relativistic nature of the Coulomb interaction, though \cite{igor}).   

To lowest order in the magnetic field 
the hydrodynamic results of Ref.\cite{hydro} reduce to the relations  
\bea
\sigma\sim\sigma_{coh}+\sigma_{in}\nonumber\\
\sigma_H\sim{B\sigma_{coh}(\sigma_{coh}+2\sigma_{in})n^{-1}}\nonumber\\
\alpha\sim{s\sigma_{coh}n^{-1}}\nonumber\\
\alpha_H\sim{sB\sigma_{coh}(\sigma_{coh}+\sigma_{in})n^{-2}}\nonumber\\
\kappa\sim{s^2T\sigma_{coh}n^{-2}}\nonumber\\
\kappa_H\sim{Bs^2T\sigma^2_{coh}n^{-3}}
\eea
Regarding these expressions the all-time important issue has been that of 
the intrinsically additive or 'inverse Matthiessen' (for the origin of this popular oxymoron
see \cite{dvk3}) structure of the kinetic coefficients.

In particular, the hydrodynamic (as well as the alternate  
memory-matrix) calculations of the DC conductivity revealed its decomposition   
onto the generalized coherent ('Drude') contribution \cite{blaise} 
\be
\sigma_{coh}={\chi^2_{JP}\over \chi_{PP}\Gamma_{mr}}
\ee
and its intrinsic 'incoherent' counterpart. In the relativistically invariant  
holographic context it was estimated as 
\be
\sigma_{inc}\sim({sT\over \epsilon+P})^2
\ee
where $s$ and $ \epsilon+P$ are entropy and enthalpy densities, respectively  \cite{herzog}.

The coherent term (21)  is controlled by the momentum relaxation rate $\Gamma_{mr}$ 
together with the momentum-momentum  $\chi_{PP}$ and 
current-momentum $\chi_{JP}$ susceptibilities. The latter   
vanishes if the operator of electric current 
is orthogonal to that of momentum. 

The coherent term receives contributions from all 
the sources of momentum relaxation (impurities, phonons, umklapp, boundary scattering, etc.).  
In turn, the second term provides for a finite conductivity in a neutral relativistic plasma in the absence of any external mechanism of momentum relaxation. Physically, it is due to the momentum-conserving Coulomb drag between the opposite charge carriers. 

Similar incoherent terms were argued to appear in the other thermo-electric 
coefficients, $\alpha_{inc}= -\mu\sigma_{inc}/T$
and ${\bar \kappa}_{inc}= \mu^2\sigma_{inc}/T$ which, however,
cancel against each other in the zero-current coefficient $\kappa$. 
 Recently, it was argued that similar terms
must be introduced into the Hall components of the kinetic coefficients 
as well \cite{amore4}. 

According to the popular scenario of Ref.\cite{blake}, in the conjectured quantum-critical regime 
the incoherent contribution $\sigma_{inc}$ is supposed to 
dominate the Ohm  conductivity, thus determining the exponent $\alpha$, 
while the Hall response would be controlled by $\sigma_{coh}$. 

Elaborating further on this proposal, in Ref.\cite{amore2} 
the coherent and incoherent terms, alongside the carrier density $n$, 
were chosen to behave as 
\bea
\sigma^{\cite{amore2}}_{coh}\sim T^{-2},~~~~~ \sigma^{\cite{amore2}}_{inc}\sim T^{-1},\nonumber\\
~~~~~n^{\cite{amore2}}\sim T^0,~~~~~s^{\cite{amore2}}\sim T
\eea
as if the fermion system was deep in the degenerate regime and 
had a well developed Fermi surface with a finite Fermi momentum $\sim n^{0.5}$.
 
Besides, the scenario of Ref.\cite{amore2} produced a list of other exponents
\bea
\gamma^{\cite{amore2}}_{hydro}=-3,~~~~~\mu^{\cite{amore2}}_{hydro}=-1, \nonumber\\ 
\lambda^{\cite{amore2}}_{hydro}=1,~~~~~\rho^{\cite{amore2}}_{hydro}=0
\eea
that could be contrasted against the data (3,4,7) as well ($spoiler$: with only a limited success).
 
Also, the assumptions (23) were made in Ref.\cite{crem}
where yet another version of the holographic, the so-called DBI, approach was utilized, 
thus resulting in the same ostensible match for the experimental Eqs.(1-3). 

However, in reality the desired dependencies (23) might be rather difficult to conform to.
Specifically, for a temperature-independent 
density $n$ the low-$T$ behavior becomes non-relativistic and 
Eq.(22) yields $\sigma_{inc}\sim s^2T^2$. 
Instead of behaving as $T^{-1}$, as per Eq.(23),
$\sigma_{inc}$ then vanishes with temperature as $T^4$ 
and, therefore, could hardly compete with $\sigma_{coh}\sim 1/\Gamma_{mr}$.
Indeed, the rate $\Gamma_{mr}$ either remains almost constant (impurity scattering) or even decreases 
with decreasing $T$ (phonons or Baber umklapp scattering). 
Either way, the assumed $T^{-1}$ behavior does not readily occur. 

In the opposite, high-$T$, limit Eq.(22) approaches a 
temperature-independent constant of order unity (or, rather, $e^2/h$). This would be typical  
for, e.g., (pseudo)relativistic $2+1$-dimensional fermions in mono-layer graphene 
which are governed by the unscreened $3$-dimensional Coulomb interactions.

Interestingly enough, this behavior would also be shared by the zero-density fermions 
with a quadratic dispersion, akin to that in (untwisted) bilayer graphene. 
In the latter case, the density of thermal excitations $n\sim T$ 
would cancel against the inelastic Coulomb scattering rate (10), thus yielding an (approximate) constant \cite{bilayer}. 

However, it should be kept in mind that in the limit of a strongly $T$-dependent carrier density, the criterion of 'hydrodynamicity' 
($\Gamma\gg\Gamma_{mr}$) further decouples from the assumed dominance of 
$\sigma_{inc}$ which would be, by and large, 
controlled by $n/\Gamma_{in}$ with the pertinent inelastic rate given by, e.g., Eq.(9) or (10). 
As a result, the range of parameters at which hydrodynamics is expected 
to work might be bounded at, both, low- and high-$T$.   

Indeed, in the presence of competing sources of momentum relaxation
such recognized hydrodynamic systems as monolayer and magic-angle-twisted bi-layer graphene   
were predicted to manifest their fluid-like behavior only within a relatively narrow window 
of temperatures where the disorder and phonon scattering mechanisms
set the lower and upper bounds, respectively. On the other hand, in the untwisted bilayer graphene
the hydrodynamic regime is not expected to be bounded from above. \cite{bilayer}.

In that regard, it is instructive to mention the recent work on the compound $BSCO$ with a low-$T_c\sim 10 K$ \cite{amore3} which 
reported the different measured exponents 
\be 
\beta^{\cite{amore3}}_{exp}=-1.5,~~~~~ \delta^{\cite{amore3}}_{exp}=-3, ~~~~~
\mu^{\cite{amore3}}_{exp}=-2.5
\ee
The new values of $\delta$ and $\mu$ were  extracted from the data taken in 
the narrow range of temperatures between $20$ and $40$ or $60$ K (above which the Nernst signal changes sign), respectively.  
For comparison, the results of Ref.\cite{griss1} for $\sigma_H, \kappa_H$, 
and $L_H$ in underdoped $YBCO$ were collected over a wider range of temperatures, yet the authors refrained from 
fitting them with any particular power-laws.

Nonetheless, the concomitant theoretical analysis of Ref.\cite{amore3} based on the hydrodynamic Eqs.(20) 
claimed to have been able to explain all of the Eqs.(1,4,25) by making the assumptions 
\bea
\sigma^{\cite{amore3}}_{coh}\sim T^0,~~~~~\sigma^{\cite{amore3}}_{inc}\sim T^{-1},\nonumber\\
n^{\cite{amore3}}\sim T^{1.5},~~~~~s^{\cite{amore3}}\sim T
\eea
Similar to the earlier proposals \cite{blake,amore2,crem} electrical transport was still going to be dominated by $\sigma_{inc}$. 
However, this time around it was supposed to occur at $low$ rather than $high$ temperatures between 
$10$ and $100$ K (despite the fact that the must-have exponent (1) could be observed up to $700$ K). 

Moreover, in Ref.\cite{amore3} the main source of momentum relaxation was attributed
to the non-quasi-particle transport through a charge density wave (CDW) \cite{del} and  
an intricate cancellation between the different 
$T$-dependent factors in $\sigma_{coh}$ was required to achieve (26).  

Notably, as compared to Ref.\cite{amore2}, 
in Ref.\cite{amore3} the definitions of $\sigma_{coh}$ and $\sigma_{inc}$ were switched 
around - presumably, to encourage the attentive readers to remain vigilant (curiously, though, 
both Refs.\cite{amore2} and \cite{amore3} feature the same first author). 
 
Lastly, by having made the above (somewhat overly flexible) assumptions, 
Ref.\cite{amore3} would also be forced into the less wanted predictions 
\be
\rho^{\cite{amore3}}_{hydro}=-1,~~~~~ \lambda^{\cite{amore3}}_{hydro}=-1.5
\ee
which are not immediately supported by the data.

In fact, if the agreement asserted in Ref.\cite{amore3} were indeed there, the assumed behavior of the carrier density (26) would have 
appeared to be in conflict with the underlying assumption of the relativistic (that is, $z=1$) kinematics of carriers,
as well as the conjectured scaling of entropy. Besides, it would also call for the inelastic 
scattering rate $\Gamma_{in}\sim n/\sigma_{inc}\sim T^{2.5}$ for which there seems to be no known microscopic mechanism. 

Indeed, in a generic $d$-dimensional system with the dispersion $\epsilon\sim p^z$
the carrier's density scales as $n\sim T^{d/z}$. Thus, taken at its face value
Eq.(26) would have implied $z=4/3$ for $d=2$.
Incidentally, this value of $z$ coincides with that proposed in Ref.\cite{hk}, despite the fact 
that instead of the novel Eqs.(25) Ref.\cite{hk} aimed at reproducing 
the 'orthodox' values of the exponents in Eqs.(1-4,7).

In light of the lingering tension between the available data and all the aforementioned proposals, it might also be of interest to
point out (albeit being at some risk of repetition) that the hydrodynamic formulas could still produce   
the exponents that are equal or close to the observed ones by adopting the 
admittedly uninteresting scenario of Ref.\cite{dvk1}. 

Namely, under the minimal assumptions  
\be
\sigma_{hydro}\sim T^{-1},~~~~~ n_{hydro}\sim T
\ee 
where no distinction is to be made 
between the 'Drude' and incoherent parts of the conductivity  
and barring any fine-tuned cancellations {\it a la} Sondheimer 
the hydrodynamic Eqs.(20) yield the following exponents
\bea
\alpha^{\cite{dvk1}}_{hydro}=-1,~~~~~\beta^{\cite{dvk1}}_{hydro}=-2, ~~~~~\gamma^{\cite{dvk1}}_{hydro}=-4,\nonumber\\ 
~~~~~\delta^{\cite{dvk1}}_{hydro}=-2, 
\nu^{\cite{dvk1}}_{hydro}=1,~~~~~ \mu^{\cite{dvk1}}_{hydro}=-2,\nonumber\\
\lambda^{\cite{dvk1}}_{hydro}=0,~~~~~\rho^{\cite{dvk1}}_{hydro}=0~~~~~
\eea
The ansatz (28) describes, e.g., a system of non-degenerate 
two-dimensional fermions with a quadratic dispersion and
generic scattering rate (10). 

Of course, the simple schemes with $T$-dependent carrier density and a single scattering rate, 
including those with $n\sim T$, have been discussed since the early days of the high-$T_c$ era \cite{linear}. 
As regards the cuprates, 
the underdoped $YBaCuO$ and $HgBaCuO$ show the presence of small electron pockets, in contrast with the large hole-like 
Fermi surface which develops in the overdoped regime above the critical doping $p^{*}$. 
This is consistent with the reports of a dramatic drop in 
the low-temperature carrier density (evaluated by the Hall number $n_H$) from 
$n_H\approx 1+p$ to $n_H\approx p$ upon crossing into the pseudogap phase \cite{griss1,chang}. 

The generic rate (9) could originate from the Baber mechanism (although the applicability of 
hydrodynamics would then be rather questionable)
whose effectiveness depends on whether or not the quasiparticle dispersion, Fermi surface topology, and spatial dimension 
conspire to provide for the comparable rates of the normal and umklapp inelastic scattering processes. 
It is believed, though, that in the cuprates, both, the multi-pocketed (in the under- and optimally-doped
cases) as well as the extended concave (in the over-doped case) hole Fermi surfaces 
might comply with the necessary conditions outlined in Ref.\cite{maslov}.\\

The matters become further complicated due to the possible onset
of CDW order, as in orthorhombic  $YBCO$ and tetragonal $HgBaCuO$  
which may induce Fermi-surface reconstruction. However, some authors believe that 
this might be a secondary phenomenon occurring only in high
magnetic fields and at temperatures below the zero-field $T_c$ \cite{pelc}. 

By contrast, when coupled with 
the Fermi liquid-like scattering rate (9) observed across much of the entire cuprates' phase diagram \cite{barisic} 
the $T$-dependent carrier density may be hinting at some alternate theoretical scenaria that neither exploit 
the notion of quantum criticality, nor attribute any special role to the incidental CDW order. In particular, 
the work of Ref.\cite{pelc} emphasizes a potential importance of the local (pseudo)gaps produced by some 
intrinsic microscopic inhomogeneity (which type of local physics would unlikely 
be conducive to any holographic speculations).\\

\noindent 
{\it Summary}\\

To summarize, this note provides another exposition of the systematic problems inherent to any approach
that is based on technical convenience, rather than physical insight. 

Of course, it can not be excluded that the seemingly suggestive scaling exhibited by a 
variety of the experimental probes  is, in fact, only approximate and 
limited to certain, insufficiently broad, ranges of parameters, thus making it virtually  
impossible to explain all such findings within the same paradigm.

Moreover, as regards the general holographic task of constructing a comprehensive catalog of all the 
different types of 'strange-metallic' behavior, the traditional focus on the cuprates 
appears to be much too narrow. 
To that end, there exists a plethora of other (e.g., heavy-fermion)  
compounds where the unexplained power-law dependencies are abound. 
As an added challenge, in many cases the apparent exponent $z$ 
remains finite, thus breaking out of the restrictive confines of the $AdS_2$ scenario characterized by the diverging $z$. 

Therefore, instead of pursuing the toxic trend of striving to reproduce  
the selected experimental plots 'at all costs', the field of applied holography 
could probably be better off using its powerful resources to focus on 
the burden of actually proving that its basic principles are sound and its computational machinery is not merely an exercise
in solving for linear perturbations about the opportunistically chosen classical gravitational backgrounds. 

Should such a proof be furnished, there should still be a plenty of other condensed matter systems \cite{abbomonte}
waiting to be tackled by 'the powerful method for studying strongly correlated systems' that 
applied holography purports to be.

\end{document}